\documentclass[showpacs,amssymb,aps]{revtex4}
\usepackage{amsmath}
\usepackage{amsfonts}
\usepackage{amssymb}
\usepackage{graphicx}
\usepackage{epsfig}
\usepackage{dcolumn}
\usepackage{bm}

\def\lessim{\lower.5ex\hbox{$\; \buildrel < \over \sim \;$}}
\def\gtrsim{\lower.5ex\hbox{$\; \buildrel > \over \sim \;$}}
\begin{document} \hbadness=10000
\topmargin -0.8cm\oddsidemargin = -0.5cm\evensidemargin = -0.5cm
\preprint{}
\title{Pion and muon production  in   $e^-,e^+,\gamma$-plasma}
\author{Inga Kuznetsova$^1$, Dietrich Habs$^2$  and Johann Rafelski$^{1,2}$}
\affiliation{$^1$Department of Physics, University of Arizona, Tucson, Arizona, 85721, USA}
\affiliation{$^2$Department f\"ur Physik der Ludwig-Maximilians-Universit\"at M\"unchen und
Maier-Leibnitz-Laboratorium, Am Coulombwall 1, 85748 Garching, Germany}
\begin{abstract}
We study production and equilibration of  pions and muons in
relativistic electron-positron-photon plasma  at
a temperature $T\ll m_\mu,\,m_\pi $. We argue that the observation of pions and muons
can be a diagnostic tool in the study of the initial properties of such a plasma formed by means of
strong  laser  fields. Conversely, properties of  muons and pions in thermal environment become
accessible to precise experimental study.
\end{abstract}

\date{May 28, 2008}

\maketitle

\section{Introduction}
The formation of a relativistic (temperature $T$  in MeV range), electron-positron-photon
  $ e^-,   e^+, \gamma $  plasma  (EP$^3$) in the laboratory using ultra-short pulse lasers is one of
the topics of current interest and forthcoming experimental effort ~\cite{TajMou,TajMouBoul}.
The elementary  properties of EP$^3$
have recently been reported, see~\cite{Thoma:2008my}, where  typical properties   are
explicitly  presented for $T=10$ MeV.
One of the challenges facing  a study of  EP$^3$ will be the understanding
of the fundamental mechanisms leading to its formation. We propose  here as a probe the production of
heavy particles with mass $m\gg  T$. Clearly, these processes occur during the history of the event
at the highest available temperature, and thus
information about the early stages of the plasma, and even pre-equilibrium state should
become accessible in this way.

We focus our attention on the strongly interacting  pions $\pi^\pm,\pi^0$ ($m_\pi c^2\lessim 140$ MeV),
and muons $\mu^\pm$($m_\mu c^2\lessim 106$ MeV), 
({\it in the following we use units in which
 $k=c=\hbar=1$ and thus  we omit these symbols from all equations.
Both, the particle mass, and  plasma  temperature, is thus given in the energy unit MeV.})
These very  heavy, compared to the electron  ($m_ec^2=0.511$ MeV), particles
are as noted natural `deep' diagnostic tools of the EP$^3$  drop. Of special interest
is  the neutral pion  $\pi^0$ which is, among all other heavy particles,
 most copiously produced for $T\ll m$.  The $\pi^0$    yield and spectrum  will
be therefore of great  interest in the study of the EP$^3$ properties.
 Conversely,   the study of  the in-medium  pion mass 
splitting  $\Delta m=m_{\pi^\pm}-m_{\pi^0}=4.594 $\,MeV    at a temperature 
$T\gtrsim \Delta m$    will contribute to the better
understanding of   this  relatively large mass splitting between
 $\pi^0$and $\pi^\pm$, $\Delta m/\overline m= 3.34\% $, believed to originate
in the isospin  symmetry breaking electromagnetic radiative corrections.

However, given its very short natural lifespan:
$$\pi^0\to \gamma+\gamma, \quad \tau_{\pi^0}^0=(8.4\pm0.6)10^{-17} {\rm s}.$$
$\pi^0$ is also the  particle most difficult to experimentally study among those we consider:
its decay products reach the detection system nearly at the same time as the electromagnetic
energy pulse of the decaying  plasma fireball, which is likely to `blind' the detectors.

This plasma drop we consider is a thousand times hotter than the center of the sun.
This implies presence of the  corresponding
high particle density $n$, energy density $\epsilon$ and pressure $P$. These quantities
 in the plasma can be  evaluated  using the relativistic
expressions:
\begin{eqnarray}
n_i &=&\int g_i f_i(p)d^3p,\\[0.3cm]
\epsilon &=& \int \sum_i g_iE_if_i(p)d^3p, \quad E_i=\sqrt{m_i^2+\vec p^{\,2}} \\[0.3cm]
P &=& \frac{1}{3}\int  \sum_i g_i\left(E_i-\frac{m_i^2}{E_i}\right)f_i(p)dp^3, \label{Ptherm}
\end{eqnarray}
where subscript $i\in \gamma$, $e^-,   e^+$, $\pi^0, \pi^+, \pi^-$, $\mu^-, \mu^+$,
$f_i(p)$ is the  momentum distribution of the particle $i$ and  and $g_i$ its degeneracy,
for $i=e^-,e^+, \gamma, \mu^-, \mu^+$ we have $g_i=2$, and $g_i=1$ for $\pi^0, \pi^- \pi^+$.
For a QED plasma which lives long enough so that electrons, positrons
are in thermal and chemical equilibrium with photons, ignoring small
QED interaction effects,  we use Fermi and
Bose momentum distribution, respectively:
\begin{equation} \label{fstat}
f_{e^\pm}  = \frac{1}{e^{(u\cdot p_{e}\pm\nu_e)/T} +1},\quad
f_{\gamma}  =  \frac{1}{e^{u\cdot p_{\gamma}/T} -1},
\end{equation}
The invariant form comprises  the Lorentz-scalar  $u \cdot p_e$,  a scalar product of
the particle 4-momentum $p^{\mu}_{i}$ with the local
4-vector of velocity $u^{\mu}$. In absence of matter flow and in the rest (in the laboratory) frame
we have
\begin{equation}
u^{\mu}=\left(1,\vec{0}\right),\label{4v}  \qquad p^{\mu}_{i}=\left(E_i, \vec{p_i}\right).
\end{equation}
When the electron chemical potential $\nu_e$ is small, $\pi T \gg \nu_e$ , the
number of particles and  antiparticles is the same,
$n_{e^-}=n_{ e^+}$. Physically, it means that the number of $e^+e^-$ pairs produced is dominating
residual matter electron yield. This is   the case for all laboratory experimental
environments of interest here, in which  $T>2 $ MeV is achieved.  We thus will set
$\nu_e=0$ in the following.

It is convenient to parametrize the electron, positron  and photon
$ e^-,   e^+, \gamma $ plasma properties   in terms of the
properties of the Stephan-Boltzmann law for  massless particles
(photons), presenting the physical properties in terms of the
effective degeneracy $g(T)$ comprising the count of all particles present at a given temperature $T$:
\begin{equation}\label{SB}
\frac{\cal E}{V}= \epsilon   =g(T)  \sigma T^4,\qquad 3P=g^\prime(T) \sigma T^4,\qquad \sigma=\frac{\pi^2}{30}.
\end{equation}
For temperatures $T\ll m_e$
we only have in this case truly massless photons and $g(T)\simeq g^\prime(T) \simeq 2_\gamma$.
Once temperature approaches and increases beyond $m_e$
we find  $g\simeq g^\prime(T) \simeq 2_\gamma+(7/8)(2_{e^-}+2_{e^+})=5.5$ degrees of freedom.
In principle these particles acquire additional in medium mass which reduces the degree of
freedom count, but this effect is compensated
by collective `plasmon' modes, thus we proceed with naive counting of nearly
free EP$^3$ components. The factor 7/8 expresses the difference in the evaluation of  Eq.\,(\ref{Ptherm})
for the  momentum distribution of  Fermions and Bosons Eq.\,(\ref{fstat}), with Bosons
 providing the reference  point at low $T$, where only massless photons are
present. In passing, we note
that in the early Universe, there  would  be further present the neutrino degrees of freedom, not
considered here for  the laboratory experiments, considering their weak coupling to matter.

\begin{figure}
\centering\hspace*{0.5cm}
\includegraphics[width=8.6cm,height=10cm]{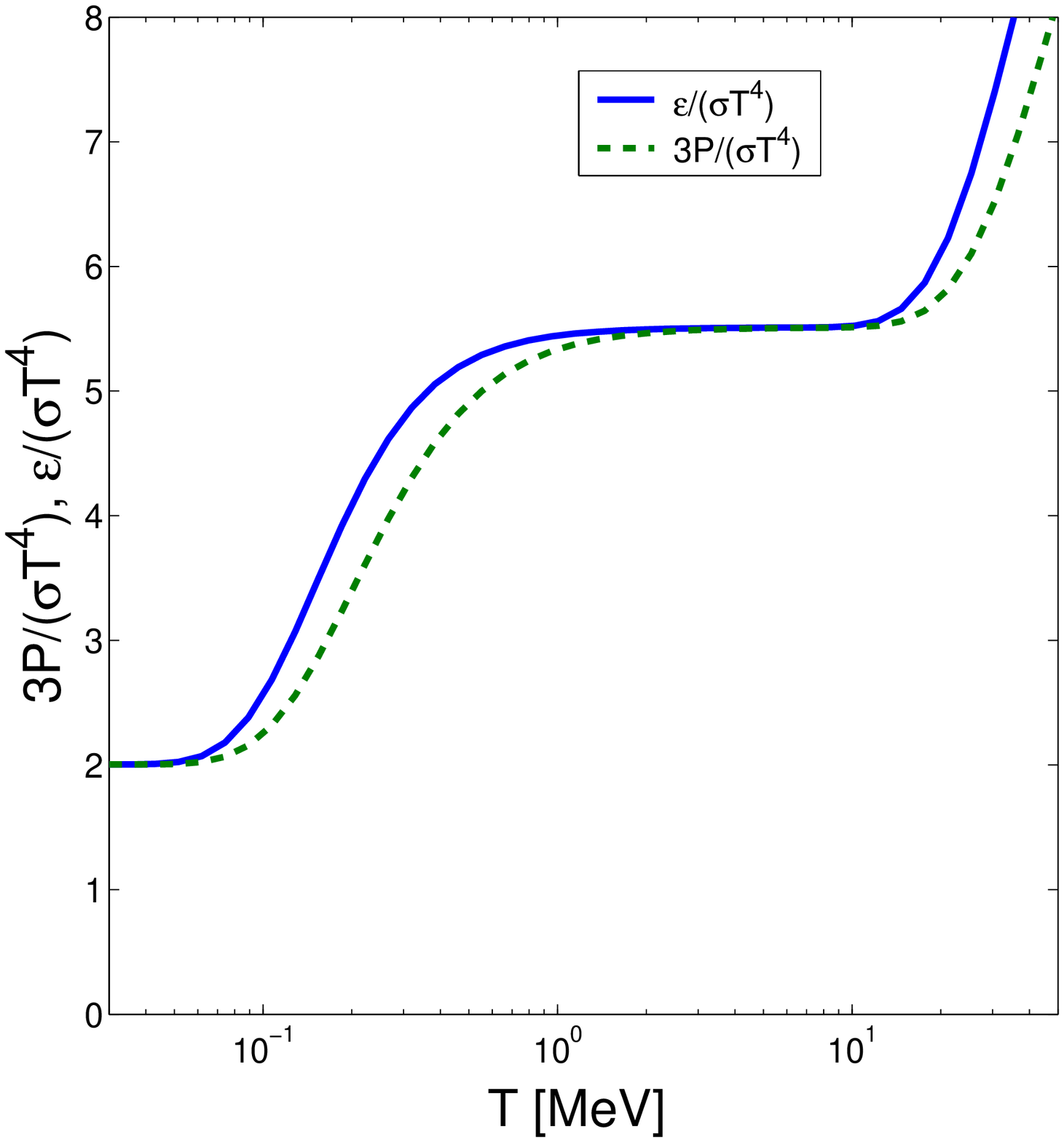}\hspace*{-0.5cm}
\includegraphics[width=8.6cm,height=10cm]{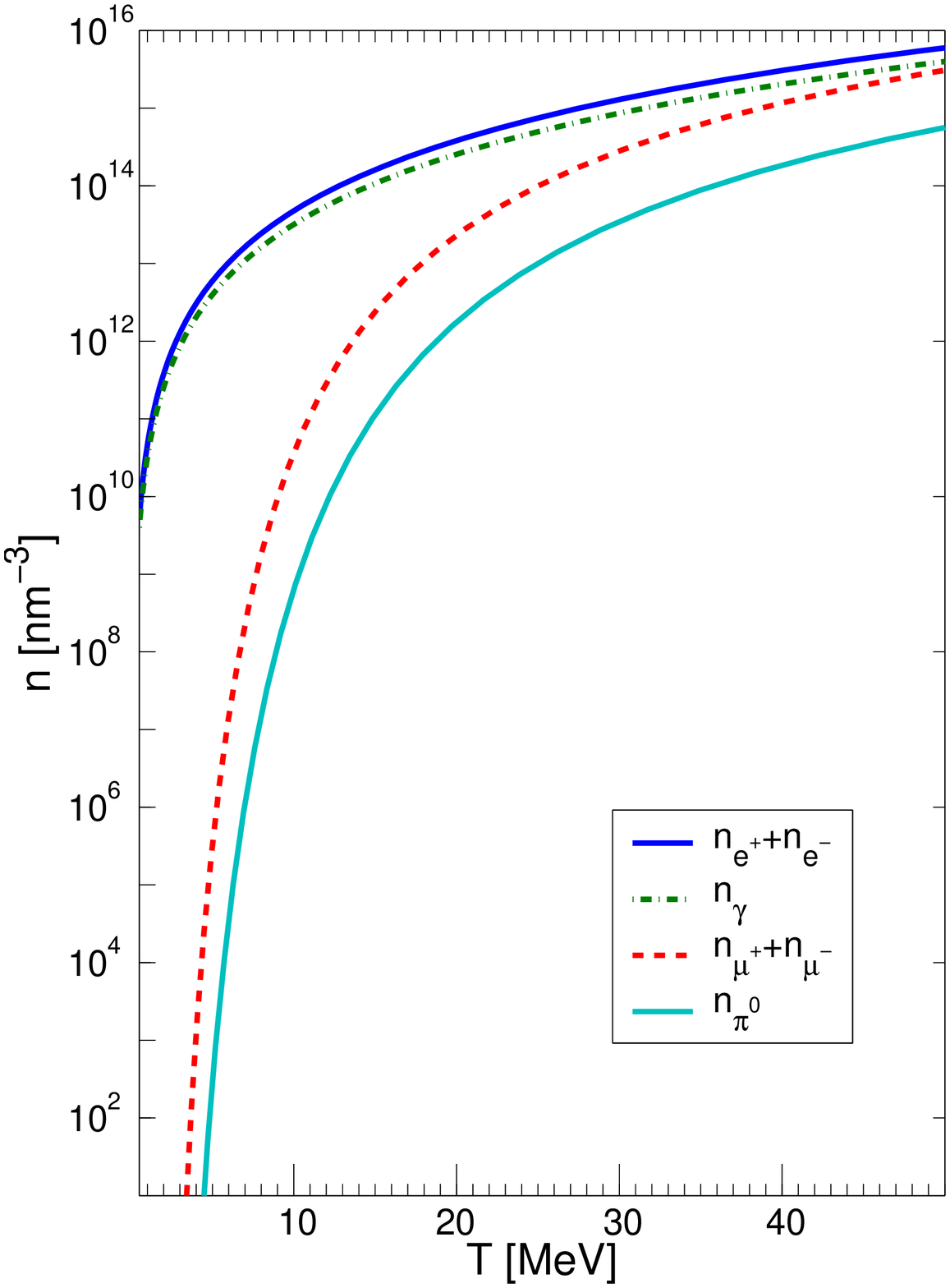}
\caption{\small{On left: the ratios $g\equiv \epsilon/\sigma T^4$ and $g'\equiv 3P/\sigma
T^4$  as a function of temperature $T$; on right: the equilibrium densities of electrons (blue, solid
line), photons (green, dash-dot line), muons (red, dashed line),
pions (blue dotted line) as functions of temperature $T$.}}
\label{energ}\label{nmupi}
\end{figure}

In figure \ref{energ} we present both $g(T)$ and  $g^\prime(T)$, as a function of temperature $T$
 in form of the energy density $\epsilon$ normalized by $\sigma T^4$, and, respectively,
the pressure $P$, normalized by $\sigma T^4/3$ .
The  $g(T)$   jumps more rapidly compared to  $g^\prime(T)$,
 between the limiting case of a black body photon
gas at  $T< 0.5$  MeV $(g=2)$ and the case $g=5.5 $ for $ \gamma$, $e^-,  e^+$, since the energy density 
also contains the rest mass energy content of all particles present.
The rise of the  ratio at $T>15$ MeV indicates  the contribution of the excitation of  muons and  pions in
equilibrated plasma. We note that the plasma  produced   pions  (and muons) are
in general  not in chemical equilibrium. The
distribution functions which maximize entropy content at given particle number and
energy content are \cite{LLStat}:
\begin{equation}
f_{\pi} = \frac{1}{\Upsilon^{-1}_{\pi^0(\pi^{\pm})}e^{u\cdot p_{\pi}/T}-1},\quad
f_{\mu} = \frac{1}{\Upsilon^{-1}_{\mu}e^{u\cdot p_{\mu}/T} +1}, \label{dfpimu}
\end{equation}
where $\Upsilon_{\pi^0(\pi^{\pm})}$ and $\Upsilon_{\mu}$ are particles fugacities.
 For $\Upsilon_i\to 0$ the quantum distributions
shown in Eq.\,(\ref{dfpimu}) turn into the classical Boltzmann distributions, with abundance prefactor $\Upsilon_i$.

In the case of interest here, when $T<m$, it suffices to consider
the Boltzmann limit of the quantum distributions Eq.(\ref{dfpimu}),
that is to drop the `one' in the denominator. Using the  the
Boltzmann momentum distribution and taking the non-relativistic
limit we have:
\begin{equation}\label{bolpio}
\frac{N_{\pi}}{V}\equiv n_{\pi} = \Upsilon_{\pi}\frac{1}{2\pi^2}Tm_\pi^2 K_2(m_{\pi}/T)
    \to \Upsilon_{\pi}  \left( \frac{m_{\pi}T}{2\pi}\right)^{3/2}e^{-m_{\pi}/T}+\ldots ,
\end{equation}
where $K_2$ (and further below also $K_1$) are the modified Bessel functions of integer order `2'
 (and `1' respectively).
 
The particle densities are shown on right in figure \ref{nmupi}. The top solid line
is the sum of $n_{e^+}+n_{e^-}$, which is marginally bigger than the
photon density (dashed, blue) which follows below.  We also include in the figure the sum density of muons
$n_{\mu^+}+n_{\mu^-}$ (red, dashed), and the density of  the neutral pion $\pi^0$ (bottom solid  line).
The chemical equilibrium corresponds to $\Upsilon_{\pi^0(\pi^{\pm})}=\Upsilon_{\mu}=1$ is used in
figure \ref{nmupi} on right, since this is the maximum density that can be reached in the buildup of
these particles, for a given temperature. 
Both heavy particle densities  appear comparatively  small in the temperature range of interest.  
However, in magnitude they rival
the normal atomic density ($\simeq 10^2/{\rm nm}^3$)already at $T=4$ MeV, and 5 MeV, respectively.
This high particle density in the chemically equilibrated plasma
  explains the relatively large collision and reaction
rates we obtain in this work. In turn, this opens the question how such dense, chemically equilibrated EP$^3$  state
can be formed -- we observe that colliding two ultra intense
circularly polarized and focused laser beams on a heavy thin metal foil(s)  is the current line of approach.
Initial simulations were performed~\cite{Shen:2001}. Many strategies can be
envisaged aiming to deposit the laser pulse energy in the smallest possible spatial and temporal volume and
 this interesting and challenging topic will without doubt keep us and others busy in years to come.

As it turns out, even a   small drop of  EP$^3$ plasma with a size scale of 1nm is, given the high particle density,
 opaque. The  mean free paths $l_i$ of  particles `i' are relatively short, at sub nano-scale~\cite{Thoma:2008my}:
\begin{equation}
l_e\simeq \left (\frac{10{\rm \,MeV}}{T}\right)^3\left (\frac{E}{ 31.1 {\rm \,MeV}}\right)^2 0.37 {\rm nm},\qquad
l_\gamma \simeq \left (\frac{10{\rm \,MeV}}{ T}\right)^2\left (\frac{E}{ 27.5 {\rm \,MeV}}\right) 0.28 {\rm nm}.
\end{equation}
Where the reference energy values (31.1 and 27.5 MeV) correspond to the mean particle energy at $T=10$ MeV.
Photons are subject to Compton scattering, and electrons and positrons to charged particle scattering. In fact these
values of $l_i$ are likely to be upper limits, since Bremsstrahlung type processes are believed to further increase opaqueness
of the plasma~\cite{ThomaPriv}.
In our considerations  plasma  particles of energy above 70 MeV are of  interest,
since these are responsible for the production
of heavy particles. We see that the mean free path of such
particles has also nm scale magnitude.

We note that a EP$^3$ drop of radius 2nm at $T=10$ MeV
contains 13 kJ energy.  This is the expected energy content of a light pulse at ELI (European Light Infrastructure, in development)
with a pulse length of about $\Delta t=10^{-14}s$. For comparison,  the maximum energy
available in particle accelerators for at least 20, if not more, years
will be in head on Pb--Pb central collisions at LHC (Large Hadron Collider) at CERN, in its LHC-ion collider mode,
where per nucleon energy of about 3 TeV is reached. Thus the total energy available is 200 $\mu$J, of which about
10\%--20\% becomes thermalized. Thus ELI will have already an overall energy advantage of $10^9$, while
in the LHI-ion case the great advantage are a) the natural localization
of the energy at the length scale of $10^{-5}$nm, given that the energy is
contained in colliding nuclei, and b) the high repetition rate of collisions.

As a purely academic exercise, we note that should one find a way
to `focus' the energy in ELI to nuclear dimensions, and scaling the energy density with $T^4$ up from
what is expected to be seen at CERN-LHC-ion ($T<1$GeV), we  exceed $T=150$ GeV, the presumed electro-weak
phase boundary. Such consideration lead the authors of Refs.~\cite{TajMou,TajMouBoul}  to suggest that
the electro-weak transition may be achieved at some future time using ultra-short laser pulses.

 Returning to present day physics, we are assuming here  that   $T$
near and in  MeV range  is achievable in foreseeable future, and that much higher values are obtainable  in
presence of pulses  with  $\Delta  t<10^{-18}s$, $c\Delta t< 0.3$nm. Hence
we consider production processes for  $\pi^0,\pi^\pm,\mu^\pm$ for $T<50$ MeV.
We   study here all two body reactions in EP$^3$ which lead to formation of the particles of interest,
excluding solely $e\gamma\to e\pi^0$, and the related $e^-e^+\to \gamma \pi^0$. The presence of a
significant (1.2\%) fraction of  $\pi^0\to e^+e^-\gamma$ decays
implies that these related two body processes could be  important in our considerations. 
However, these    reactions involve  the $\pi_0$ off-mass shell   couping to two photons, which
needs to be better understood before we can consider these reactions in our context. 

We also do not consider here the inverse three body reactions 
$ e^+e^-\gamma\to \pi^0$,  since  there is no exponential gain in using
$n>2$ particles to overcome an energy threshold, here $m_{\pi^0}$. 
The independent probability of  finding $n$ particles with energy 
$m_{\pi^0}/n$ each is the same for any value of $n$: 
\begin{equation}
P_1P_2....P_n \propto \left( e^{-m_{\pi^0}/nT}\right)^n=e^{-\frac{m_{\pi^0}}{T}}.
\end{equation}
This resolves the   argument  that more particles could  overcome 
more easily the reaction barrier. $n$-body reactions with $n>2$ 
are in fact suppressed in  EP$^3$  by the weakness of the 
electromagnetic (EM)  interaction, since adding an EM-interacting 
particle to the reactions process  requires  an EM-vertex with 
$\alpha=1/137$. Thus microscopic reactions in  EP$^3$
 involving $n>2$ are suppressed by a factor 100 
for each additional EM particle involved in the reaction.  
This does not mean that a collective/coherent  process of heavy 
particle production by many particles is similarly suppressed: 
for example fast time varying electromagnetic fields provide through 
$\vec E\cdot \vec B$ a collective source of $\pi^0$. We defer further 
study of this production mechanism which requires 
 multi MeV$^{-1}$ range oscillation  to be present in EP$^3$.

In the following section, we introduce the master equation governing
the production of pions and muons in plasma and formulate
the invariant rates in terms of know physical reactions. In section \ref{resnum}
we obtain the  numerical results for particles production rates and
reactions relaxation times which we present as figures.
In section \ref{concl} we  discuss these results further and consider
their implications.

\section{Particles production rates} \label{master}
\subsection{$\pi^0$ production}
$\pi^0$ in the QED plasma is  produced
predominantly  in the thermal two photon fusion~\cite{KuznKodRafl:2008}:
\begin{equation}
\gamma+\gamma \rightarrow \pi^0. \label{pi0gg}
\end{equation}
Much less probable is the production of $\pi_0$ in the reaction:
\begin{equation}
e^-+e^+ \rightarrow \pi^0. \label{pi0ee}
\end{equation}
These formation
processes are the inverse of the decay process of $\pi_0$.
The smallness of the electro-formation of $\pi_0$ is characterized by
the small  branching ratio in $\pi_0$ decay
$B=\Gamma_{ee}/\Gamma_{\gamma\gamma}=6.2\pm 0.5 10^{-8}$.
Other  decay processes involve more than two particles.  
$\pi^0$ can also be formed by charged pions in charge exchange
reactions. However,  in EP$^3$  in the domain of $T$ of interest
we find that at first the neutral pions will be produced. These
in turn produce charged pions. Therefore we introduce the pion charge exchange
process in the context  of charged pion formation  in  the subsection \ref{pichprod}, 
and since these can be important, we show these explicitely here as well.

Omitting all sub-dominant   processes, the resulting master equation for  neutral  pion  number evolution is:
\begin{eqnarray}
 \frac{1}{V}\frac{dN_{\pi^0}}{dt}&=& \frac{d^4W_{\gamma \gamma \rightarrow\pi^0}}{dVdt} 
  -\frac{d^4W_{\pi^0 \rightarrow \gamma \gamma}}{dVdt}\nonumber\\
&+&\frac{d^4W_{\pi^+ \pi^- \rightarrow\pi^0+\pi^0}}{dVdt} 
  -\frac{d^4W_{\pi^0 +\pi^0\rightarrow \pi^+ \pi^- }}{dVdt},
\label{piev}
\end{eqnarray}
where $N_{\pi^0}$ is total number of $\pi^0$, $V$ is volume
of the system, ${d^4W_{\gamma \gamma \rightarrow\pi^0}}/{dVdt}$ 
 is the (Lorentz) invariant $\pi^0$ production rate per unit time and volume in photon  fusion, and
$d^4W_{\pi^0 \rightarrow \gamma \gamma}/dVdt$ is the  invariant
$\pi^0$ decay rate per unit volume and time. 
Similarly, ${d^4W_{\pi^+ \pi^- \rightarrow\pi^0+\pi^0}}/{dVdt} $ is the  pion charge 
exchange  $\pi^0$ production rate per unit time and volume
while ${d^4W_{\pi^0 +\pi^0\rightarrow \pi^+ \pi^- }}/{dVdt} $ is the   
corresponding  reverse  reaction loss rate.

We assume that in the laboratory frame the momentum
distribution of produced $\pi^0$ are characterized
by the ambient temperature.
Eq.\,(\ref{bolpio})   defines the relation of fugacity $\Upsilon_{\pi} $ to the
yield. This equation allows now to   study the production dynamics as if
we were dealing with a  $\pi^0$ in a thermal bath, and to exploit
the detailed balance between decay and production process in order
to estimate the rate of  $\pi^0$ production. This theoretical  consideration
should not be understood as assumption of equilibration of $\pi^0$,
which could upon production escape from the small plasma
drop.

In \cite{KuznKodRafl:2008} the detailed balance relation is derived in detail,  which takes the form
\begin{equation}\label{Wrate}
\Upsilon_{\pi^0}^{-1}\frac{d^4W_{\pi^0 \rightarrow \gamma\gamma}}{dVdt}=
\Upsilon^{-2}_{\gamma}\frac{d^4W_{\gamma \gamma \rightarrow \pi^0}}{dVdt}\equiv R_{\pi^0}.
\end{equation}
This allows that  Eq.(\ref{piev}) can be written in the form:
\begin{equation}
\frac{1}{V}\frac{dN_{\pi^0}}{dt} = (\Upsilon_{\gamma}^2-{\Upsilon_{\pi^0}})R_{\pi^0 }
-({\Upsilon_{\pi^0}^2}-{\Upsilon_{\pi^{\pm}}^2}){R_{\pi^0\pi^0\leftrightarrow \pi^+\pi^-}} .
\label{pieqdyn}
\end{equation}
For $\Upsilon_{\pi_0} \to \Upsilon^2_{\gamma} \to \Upsilon_{\pi^{\pm}}^2=1$  we reach
chemical equilibrium, the time variation of density due to production and decay vanishes.

The charge exchange process rate ($R_{\pi^0\pi^0\leftrightarrow \pi^+\pi^-}$, last in Eq.\,(\ref{pieqdyn}))  
balances the first contribution in Eq.\,(\ref{pisc}), where it will be further discussed. 
The rate  $R_{\pi^0} $  can be written as
\begin{eqnarray}
R_{ \pi^0}&=&
   \int\frac{d^{3}{p_{\pi}}}{(2\pi\ )^32E_{\pi}}
   \int\frac{d^{3} {p_{2\,\gamma}}}{(2\pi\ )^32E_{2\,\gamma}}
   \int\frac{d^{3}{p_{1\,\gamma}}} {(2\pi\ )^32E_{1\,\gamma} }\left(2\pi\right)^{4}
 \delta^{4}\left(p_{1\,\gamma}+p_{2\,\gamma}-p_{\pi}\right)\times \nonumber\\ &&
  \sum_{spin}\left|\langle p_{1\,\gamma}p_{2\,\gamma}\left| M\right|p_{\pi}\rangle\right|^{2}
   f_{\pi}(p_{\pi})f_{\gamma}(p_{1\,\gamma})f_{\gamma}(p_{2\,\gamma})
 \Upsilon^{-2}_{\gamma}\Upsilon_{\pi^{0}}^{-1}e^{u \cdot p_{\pi}/T}. \label{pi0pr}
\end{eqnarray}
where for $\pi^0$ formation there was the  factor $(1+f_{\pi})$ which we reduced  using the relation
\begin{equation} \label{FBrel}
 1\mp f_\pm =\Upsilon_i^{-1}e^{u \cdot p_{i}/T}f_\pm ,
\end{equation}
where Fermi $(f_+)$ and Bose $(f_-)$ distributions are implied for  particle $i$. Similarly, in the $\pi^0$-decay case 
we replaced the two stimulated decay factors $(1+f_\gamma)^2$ in that way. Eq.\,(\ref{pi0pr}) follows
Including in Eq.\,(\ref{pi0pr})  the prefactors required
by Eq.\,(\ref{Wrate}) and recalling  time reversal invariance, i.e. $M=M^\dagger$:
\begin{equation}\label{unitary}
\left|\langle p_{1\,\gamma}p_{2\,\gamma}\left| M \right|p_{\pi}\rangle\right|^{2}
=\left|\langle  p_{\pi} \left| M \right|p_{1\,\gamma}p_{2\,\gamma} \rangle\right|^{2} .
\end{equation}
We realize that the result, Eq.\,(\ref{pi0pr})  is manifestly symmetric for the two reaction directions. 
It is interesting to note that in Boltzmann limit all fugacities cancel in  Eq.\,(\ref{pi0pr}). 
 
We introduce  the
pion equilibration  (relaxation) time constant by:
\begin{equation}\label{taupi0}
\tau_{\pi^0}  =  \frac{{dn_{\pi^0}}/{d\Upsilon_{\pi^0}}}{ R_{\pi^0} } .
\end{equation}
Note that when the volume does not change in time on scale of $ \tau_{\pi^0} $ (absence of expansion
dilution) and thus $T$ is constant, the left hand side of Eq.(\ref{pieqdyn}) becomes
$dn_{\pi^0}/dt$. Given the relaxation time definition
Eq.(\ref{taupi0}) the time evolution  for of the pion fugacity for a
system  at  fixed time independent temperature   satisfies:
\begin{equation}\label{pidynfug}
 \tau_{\pi^0} \frac{d \Upsilon_{\pi^0}}{dt}=\Upsilon^2_{\gamma}-\Upsilon_{\pi^0}
-({\Upsilon_{\pi^0}^2}-{\Upsilon_{\pi^{\pm}}^2})\frac{R_{\pi^0\pi^0\leftrightarrow \pi^+\pi^-}}{R_{\pi^0 }}.
 \end{equation}
When the charge exchange reaction can be ignored, 
 for $\Upsilon_{\pi^0}(t=0)=0$ we find the analytical solution $\Upsilon_{\pi^0}
=\Upsilon^2_{\gamma}\left(1-e^{-t/\tau_{\pi^0}}\right)$, justifying the proposed  definition of
the relaxation constant.

We note that Eq.(\ref{pidynfug}) also
describes the   decay of a $\pi^0$. Therefore, up to small
modifications introduced  by the thermal
medium (see discussion below),
$$ \tau_{\pi^0} \simeq  \tau_{\pi^0}^0.$$
The $\pi^0$ production rate is thus related to the decay rate $1/\tau_{\pi^0}^0$ by the simple formula
\begin{equation}
R_{ \pi^0} \simeq
    \frac{{dn_{\pi^0}}/{d\Upsilon_{\pi^0}}}
{\tau_{\pi^0}^0 }\simeq \left(\frac{m_{\pi}T}{2\pi}\right)^{3/2}\frac{e^{-m_{\pi}/T} }{ \tau_{\pi^0}^0},
\label{taupi01}
\end{equation}
where in the last expression we have used Eq.\,(\ref{bolpio}) in the limit $m>>T$. It is important for
the reader to remember that   derivation of  Eq\,(\ref{taupi01}) is based on detailed balance in thermally
equilibrated plasma, and does not require chemical equilibrium to be established.

\begin{figure}
\centering
\includegraphics[width=8.6cm,height=8.5cm]{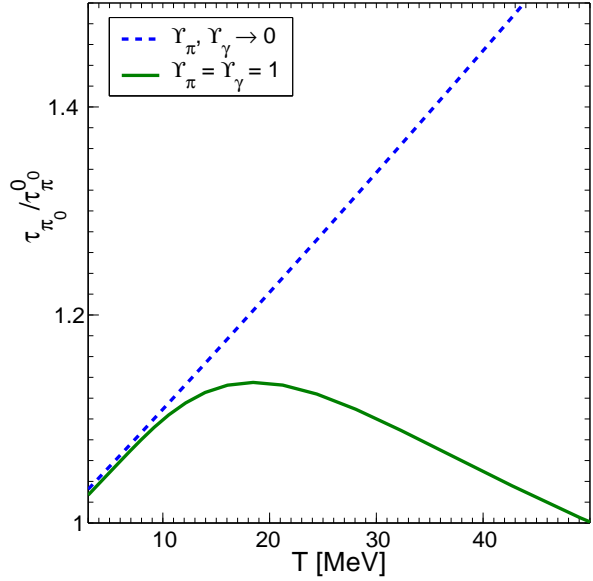}
\caption{\small{The ratios $\tau_{\pi^0}/\tau^0_{\pi_0}$ as
functions of temperature $T$ for relativistic Boltzmann limit  (blue, dashed
line) and for quantum distribution in chemical equilibrium, $\Upsilon_{\pi}=\Upsilon_{\gamma}=1$
(green, solid line).}} \label{taupi0app}
\end{figure}

Now we consider how and why $\tau_{\pi^0}\simeq \tau^0_{\pi_0}$. It turns out that there are
both relativistic and quantum effects which contribute and they (nearly) cancel:
 the relativistic effect arises because $\tau_{\pi^0}$ in Eq.(\ref{taupi01}) is in lab frame while the
known $\tau^0_{\pi^0}$ is in the pion rest frame. In the relativistic
Boltzmann limit the correction is obtained considering the related time dilation effect~\cite{KuznKodRafl:2008} is:
\begin{equation}
\tau_{\pi^0} = \frac{\tau_{\pi^0}^0}{<{1}/{\gamma}>}=\tau_{\pi^0}^0\frac{K_2(m_{\pi^0}/T)}{K_1(m_{\pi^0}/T)},
\label{taurel}
\end{equation}
where $<{1}/{\gamma}>$ is average inverse Lorentz factor. We find that this effect implies that
$\tau_{\pi^0}$ in the lab frame increases with temperature. This effect is shown by dashed
 (blue) line   in  figure \ref{taupi0app}. Furthermore,
with increasing  temperature quantum distribution functions for photons and   for
the  produced particle need to be considered. This leads to the result shown as solid line
 (green)  in  figure \ref{taupi0app}.
Thus in general $ \tau_{\pi^0} >  \tau_{\pi^0}^0$, by up to 14\%.

\begin{figure}
\centering
\includegraphics[width=8.6cm,height=8.5cm]{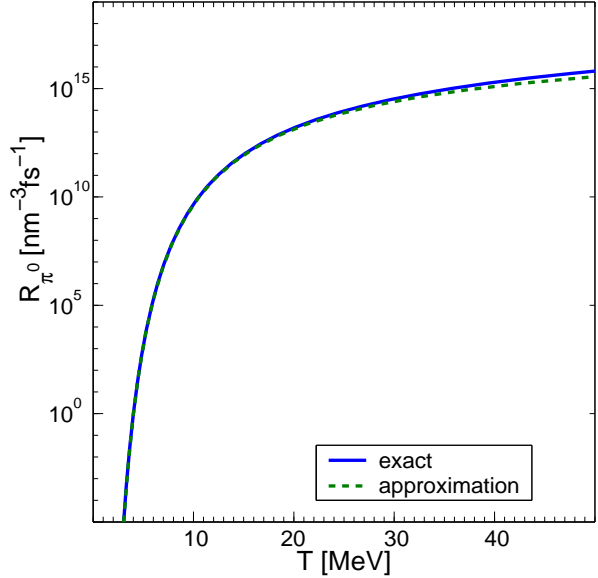}
\caption{\small{The $\pi_0$ production rate (blue, solid line)
and approximate rate from Eq.(\ref{taupi01}) (green dashed line)
as functions of temperature $T$.}} \label{Rpi0}
\end{figure}

We can further evaluate exactly the reaction rate Eq.\,(\ref{pi0pr})~\cite{KuznKodRafl:2008}:
\begin{equation}
{R_{\pi^0}} = \frac{1}{\left(2\pi\right)^{2}}\frac{m_{\pi}}{\tau_{\pi^0}^0}\int
_{0}^{\infty}\frac{p_{\pi}^{2}dp_{\pi}}{E_{\pi}}\frac{\Upsilon_{\pi^0}^{-1}e^{E_{\pi}/T}}
{\Upsilon_{\pi^0}^{-1}e^{E_{\pi}/T} - 1}\Phi(p_{\pi}),
\label{Decay1}
\end{equation}
where
\begin{equation}
\Phi\left(p_{\pi}\right)  = \int_{-1}^{1}d\zeta\Upsilon^{-2}_{\gamma}
\frac{1}{\Upsilon^{-1}_{\gamma}e^{\left( a - b\zeta\right)} - 1}
\frac{1}{\Upsilon^{-1}_{\gamma}e^{\left(a + b\zeta\right)} - 1},
\label{phif}
\end{equation}
with
\begin{equation}
a  =  \frac{\sqrt{m_{\pi}^2+p_{\pi}^2}}{2T};  \quad
b  = \frac{p_{\pi}}{2T}.
\end{equation}
This integral for $\Upsilon_{\gamma}=1$ takes the form:
\begin{equation}
\Phi(p_{\pi^0})=\frac{2}{b(e^{2a} -  1)}\left(b+\ln\left(1+\frac{\left(e^{(b-a)}-e^{-(a+b)}\right)}
{\left(1-e^{b-a}\right)}\right) \right). \label{phiab}
\end{equation}
This exact result (blue, solid line) is compared to the approximate result  Eq.(\ref{taupi01}) (green, dashed line)  in figure  \ref{Rpi0}.
We note that it is hard to discern a difference on logarithmic scale, especially so for small temperatures where  the only (small) effect is
the relativistic time dilation. This implies that it is appropriate to use the simple  result  Eq.(\ref{taupi01}) in the study of $\pi^0$ production.

Before closing this section we note that we can use exactly the same method to extract from the partial
width of the $\pi_0\to e^+e^-$ the reaction rate for the inverse process, which will be discussed below. All arguments carry
through in identical and exact fashion  replacing where appropriate the Bose by Fermi distributions and using Eq.\ref{FBrel}.

\subsection{Muon  production}\label{muprod}
In the plasma under consideration, muons can be directly  produced in the reactions:
\begin{eqnarray}
\gamma + \gamma \rightarrow \mu^{+} + \mu^{-}, \label{ggmu}\\
e^{+} + e^{-} \rightarrow \mu^{+} + \mu^{-}. \label{eemu}
\end{eqnarray}
For reactions (\ref{ggmu}) and (\ref{eemu}) the master evolution equation developed for the study of thermal
strangeness in heavy ion collisions applies
~\cite{Biro:1981zi,Rafelski:1982pu,Matsui:1985eu,Koch:1986ud} (compared to these references our 
definition is changed, their  $R_{12 \rightarrow 34}\rightarrow 
R_{12 \rightarrow 34}/(\Upsilon_1\Upsilon_2)$ in order to make the forward-backward symmetry explicit )
\begin{equation}
\frac{1}{V}\frac{dN_{\mu}}{dt} = 
(\Upsilon_{\gamma}^2 -\Upsilon_{\mu}^2){R_{\gamma\gamma \leftrightarrow \mu^+\mu^-}} +
(\Upsilon_e^2 -\Upsilon_{\mu}^2)R_{e^+e^- \leftrightarrow \mu^+\mu^-}. \label{muev}
\end{equation}
Like before for $\pi^0$ we consider the master equation in order to find appropriate definition of the
relaxation time constant for $\mu^\pm$ production. In no way should this be understood to imply
that muons are retained in the small plasma drop.  In chemically equilibrated EP$^3$
the $\mu$ production relaxation time is defined by:
\begin{equation}\label{taumu}
\tau_{\mu} =
\frac{1}{a}\frac{dn_{\mu}/d\Upsilon_{\mu}}{\left(R_{\gamma\gamma\leftrightarrow \mu^+\mu^-} + 
               R_{e^+e^- \leftrightarrow \mu^+\mu^-}\right)},
\end{equation}
where a suitable choice is $a=1,2$ for $t=0,\infty$, respectively (see below).
The form of Eq.\,(\ref{taumu})  assures that, omitting the volume expansion, i.e. the dilution
effect,   the evolution of the muon fugacity obeys the equation
\begin{equation}
a\tau_{\mu}  \frac{d\Upsilon_{\mu}}{dt}=1-\Upsilon_{\mu}^2,\quad \Upsilon_{\gamma}=\Upsilon_{e}=1,
\end{equation}
which  has for $\Upsilon_\mu(t=0)=0$ the simple analytical solution~\cite{Rafelski:1982pu}:
  \begin{equation}
\Upsilon_{\mu}=\tanh t/a\tau_\mu .
\end{equation}
For $t\to \infty$, near to chemical equilibrium,  $\Upsilon_{\mu}\to 1-e^{-2t/a\tau_\mu}$, while
  for $t\to 0$, at the onset of particle production with small $\Upsilon_{\mu}$ we have
$\Upsilon_{\mu} ={t/(a\tau_\mu)}$ . Hence,  near to chemical equilibrium  it
is appropriate to use $a=2$ in definition of relaxation time Eq.(\ref{taumu}),
while at the onset of particle production, more
applicable to this work a more physical choice would be $a=1$. However,  following the convention, in
the results presented below the value $a=2$ is used.

The invariant muon  production rate in photon fusion as introduced above is:
\begin{eqnarray}
 R_{\gamma \gamma \leftrightarrow \mu^+\mu^-}&=& \int\frac{d^{3}{p_{\mu^+}}}{(2\pi)^32E_{\mu^+}}
    \int\frac{d^{3}{p_{\mu^-}}}{(2\pi)^32E_{\mu^-}}\int\frac{d^{3}{p_{1\,\gamma}}}{(2\pi)^32E_{1\,\gamma} }
\int\frac{d^{3}{p_{2\,\gamma}}}{(2\pi)^32E_{2\,\gamma} }\left(2\pi\right)^{4}
\delta^{4}\left(p_{1\,\gamma}+p_{2\,\gamma}-p_{\mu^+}-p_{\mu^{-}}\right)\times\nonumber\\
&&\hspace*{-0.8cm}\sum_{\rm spin}\left|\langle p_{1\,\gamma}p_{2\,\gamma}\left|
M_{\gamma\gamma\rightarrow \mu^+\mu^-}\right|p_{\mu^+}p_{\mu^-}\rangle\right|^{2}
 f_{\gamma}(p_{1\,\gamma})f_{\gamma}(p_{2\,\gamma})f_{\mu}(p_{\mu^+})f_{\mu}(p_{\mu^-})
\Upsilon_{\gamma}^{-2}\Upsilon_{\mu}^{-2}e^{u \cdot (p_{\mu^+} + p_{\mu^-})/T}\label{ggmumu}
\end{eqnarray}
and the invariant muon production rate in electron-positron fusion is:
\begin{eqnarray}
R_{e^+ e^- \leftrightarrow \mu^+\mu^-}&=&
\int\frac{d^{3}{p_{\mu^+}}}{(2\pi)^3 2E_{\mu^+}}
\int\frac{d^{3}{p_{\mu^- }}}{(2\pi)^3 2E_{\mu^- }}
\int\frac{d^{3}{p_{e^+}}} {(2\pi)^32E_{e^+}}
\int\frac{d^{3}{p_{e^- }}} {(2\pi)^3 2E_{e^-}}
\left(2\pi\right)^{4}\delta^{4}\left(p_{e^+}+p_{e^-}-p_{\mu^+}-p_{\mu^{-}}\right)\times \nonumber\\
&&\hspace*{-0.8cm}\sum_{\rm spin}\left|\langle p_{e^+}p_{e^-}\left|
M_{e^+e^-\rightarrow \mu^+\mu^-}\right|p_{\mu^+}p_{\mu^-}\rangle\right|^{2}
f_{e}(p_{e^+})f_{e}(p_{e^-})f_{\mu}(p_{\mu^+})f_{\mu}(p_{\mu^-})
\Upsilon_{e}^{-2}\Upsilon_{\mu}^{-2}e^{u \cdot (p_{\mu^+} + p_{\mu^-})/T}.\label{eemumu}
\end{eqnarray}
We note that  in Eq.\,(\ref{ggmumu}) and Eq.\,(\ref{eemumu})  in the Boltzmann limit all fugacities cancel, and that 
the forward-backward reaction symmetry is explicit. Moreover, it is interesting to note that despite
inclusion of quantum effects (Bose stimulated emission and/or Fermi blocking), when using rates as defined in this paper,
we don't change the master population equation system arising for Boltzmann particles. 
The only modification is a slight fugacity dependence 
of rates presented in Eq.\,(\ref{pi0pr}), Eq.\,(\ref{ggmumu}), Eq.\,(\ref{eemumu}).

The $\sum |M_{e^+e^-\rightarrow \mu^+\mu^-}|^2$ differs from often considered heavy quark production
$\sum |M_{q\bar{q}\rightarrow c\bar{c}}|^2$~\cite{Combridge:1978kx, Gluck:1977zm}
($m_c>>m_q$) by color factor $2/9$, and the coupling
$\alpha_s\to \alpha $ of QCD has to be changed to   $\alpha=1/137$ of QED. Then we obtain, based on above references:
\begin{equation}
\sum |M_{e^+e^-\rightarrow \mu^+\mu^-}|^2=g_e^28\pi^2\alpha^2\frac{(m^2-t)^2+(m^2-u)^2+2m^2s}{s^2}, \label{m2emu}
\end{equation}
where $m=106$ MeV is the muon mass, electron and positron degeneracy $g_e=2$, and  $s$, $t$, $u$ are the usual
Mandelstam variables: $s=(p_{1}+p_{2})^2$, $t=(p_{3}-p_{1})^2$, $u=(p_{3}-p_{2})^2$, $s+t+u=2m^2$.
For the total averaged over initial states $|M|^2$ for photon fusion we have
\begin{equation}
|M_{\gamma\gamma \rightarrow \mu^+\mu^-}|^2
 =g_\gamma^2 8 \pi^2 \alpha^2 \left(-4\left(\frac{m^2}{m^2-t}
 +\frac{m^2}{m^2-u}\right)^2+4\left(\frac{m^2}{m^2-t}+\frac{m^2}{m^2-u}\right)
+\frac{m^2-u}{m^2-t} + \frac{m^2-t}{m^2-u}\right), \label{m2gmu}
\end{equation}
where degeneracy $g_{\gamma}=2$.
Near threshold   $s \approx 4m^2$, with  $t, u  \approx - m^2$ we find
\begin{equation}
|M_{\gamma\gamma \rightarrow \mu^+\mu^-}|^2 = 64 \pi^2 \alpha^2,
\qquad
|M_{e^+e^-\rightarrow \mu^+\mu^-}|^2 = 32 \pi^2 \alpha^2.
\end{equation}
The $e^+e^-\to \mu^+\mu^- $ reaction  involves a single photon, and
thus  it is more constrained (by factor 2) compared to the photon
fusion, which is governed by two Compton type Feynman diagrams.
However, in the rate we compute below, the indistinguishability of
the two photons introduces an additional factor $1/2$, so that both
reactions differ only by the difference in the quantum Bose and
Fermi distributions.

Integrals in Eq.(\ref{ggmumu}) and (\ref{eemumu}) can be evaluated in spherical coordinates. We define:
\begin{equation}
q=p_1+p_2;\,\,\,\,p=\frac{1}{2}(p_1-p_2);\,\,\,\,q^{\prime}=p_3+p_4;\,\,\,\,p^{\prime}=\frac{1}{2}(p_3-p_4);
\end{equation}
z-axis is chosen in the direction of
$\overrightarrow{q}=\overrightarrow{p_1}+\overrightarrow{p_2}$:
$$q_{\mu}=(q_0,0,0,0),\,\,\,\,p_{\mu}=(p_0, p\sin\theta,0, p\cos\theta),\,\,\,\,
p_{\mu}^{\prime}=(p^{\prime}_0, p^{\prime}\sin\phi\sin\chi, p^{\prime}\sin\phi\cos\chi, p^{\prime}\cos\phi).$$
Now we obtain~\cite{Matsui:1985eu}:
\begin{eqnarray}
&& {R_{e^+ e^-(\gamma\gamma) \leftrightarrow \mu^+\mu^-}}= \frac{1}{1+I}\frac{(4\pi)(2\pi)}{(2\pi)^4 16} \int_{2m_{\mu}}^{\infty}
dq_0 \int_0^{s-q_0^2}dq\int_{-\frac{q}{2}}^{\frac{q}{2}}dp_0\int_{-\frac{q^*}{2}}^{\frac{q^*}{2}}dp{\prime}_0
\int_0^{\infty}dp \int_0^{\infty}dp\prime\int^{1}_{-1}d(\cos{\theta})\int^{1}_{-1}d(\cos{\phi})\
\nonumber\\[0.4cm]
&&\times\int_0^{2\pi}d{\chi}\delta\left(p-\left(p_0^2+\frac{s}{4}\right)^{1/2}\right)
 \delta\left(p{\prime}-\left(p{\prime}_0^2-{m_{\mu}^2}+\frac{s}{4}\right)^{1/2}\right)\delta \left(\cos{\theta}-\frac{q_0p_0}{qp}\right)\delta
\left(\cos{\phi}-\frac{q{\prime}_0p{\prime}_0}{qp}\right)\nonumber\\[0.4cm]
&&\times\sum|M_{e+e-(\gamma\gamma)\rightarrow \mu\mu}|^2
\Upsilon_{\mu}^{-2}f_{\mu}\left(\frac{q_0}{2}+p_0\right)f_{\mu}\left(\frac{q_0}{2}-p_0\right)
\Upsilon_{e(\gamma)}^{-2}f_{e(\gamma)}\left(\frac{q_0}{2}+p^{\prime}_0\right)f_{e(\gamma)}
\left(\frac{q_0}{2}-p^{\prime}_0\right)\exp{(q_0/T)}, \label{ratemu}
\end{eqnarray}
where
$q^*={q}\sqrt{1-\frac{m_{\mu}^2}{s}}$. The integration over $p$,
$p^{\prime}$, $\cos{\theta}$, $\cos{\phi}$ can be done analytically
considering the delta-functions. The other integrals can be evaluated
numerically. For the case of indistinguishable  colliding particles (two photons)   there is additional factor $1/2$
implemented by the value $I=1$, while for distinguishable colliding particles (here electron and positron) $I=0$.

\subsection{$\pi^{\pm}$ production}\label{pichprod}
$\pi^{\pm}$ can be produced in $\pi_0\pi_0$ charge exchange scattering:
\begin{equation}
\pi^0 + \pi^0 \rightarrow \pi^{+} + \pi^{-}, \label{pppp}
\end{equation}
as well as  in two photon, and  in electron-positron fusion processes
\begin{eqnarray}
\gamma+\gamma \rightarrow \pi^{+} + \pi^{-}, \label{ffpp} \\[0.2cm]
e^+ + e^- \rightarrow \pi^{+} + \pi^{-}.  \label{eepp}
\end{eqnarray}
We find  that for $\pi^{\pm}$ production, the last two processes are much
slower compared to the   first,  in case that $\pi_0$ density is near chemical equilibrium.
Similarly, the two photon fusion  to two $\pi^0$:
\begin{equation}
\gamma + \gamma \rightarrow \pi^0 + \pi^0, \label{ggpi0pi0}
\end{equation}
turns out, as expected, to be much smaller than one $\pi^0$ production. It is
a reaction of higher order in $\alpha$ and the energy is shared between two
final particles.

The time evolution equations for the number
of $\pi^{\pm}$ are similar to Eq. (\ref{muev}):
\begin{eqnarray}
\frac{1}{V}\frac{dN_{\pi^{\pm}}}{dt} =
({\Upsilon_{\pi^0}^2}-{\Upsilon_{\pi^{\pm}}^2}){R_{\pi^0\pi^0\leftrightarrow \pi^+\pi^-}} 
+ (\Upsilon^2_{\gamma} - {\Upsilon_{\pi^{\pm}}^2}){R_{\gamma\gamma \leftrightarrow\pi^+\pi^-}}
+(\Upsilon_e^2 - {\Upsilon_{\pi^{\pm}}^2}){R_{e^+e^- \leftrightarrow\pi^+\pi^-}}. \label{pisc}
\end{eqnarray}
In order to evaluate the pion  production rates in two body processes  we use reaction cross
 section, and the relation~\cite{Letessier:2002gp}:
\begin{equation}
{R_{1\,2 \leftrightarrow \pi^+\pi^-}} = \frac{g_1g_2}{32\pi^4}\frac{T}{1+I}
\int_{s_{th}}^{\infty}ds\sigma(s)\frac{\lambda_2(s)}{\sqrt{s}}K_1(\sqrt{s}/T),
\end{equation}
(compared to reference~\cite{Letessier:2002gp} our definition is changed 
$R_{12\rightarrow 34} \rightarrow 
R_{12 \rightarrow 34}/(\Upsilon_1 \Upsilon_2)$)
where
\begin{equation}
\lambda_2(s) = (s-(m_{1}+m_2)^2)(s-(m_1-m_2)^2),
\end{equation}
$m_1$ and $m_2$, $g_1$ and $g_2$, $\Upsilon_1$ and $\Upsilon_2$ are masses, degeneracy and fugacities of initial interacting particles.

For the respective three cross sections we use, all results valid in the common range  $s\le 1 $ GeV$^2$ we consider here:
\begin{itemize}
\item
The  cross section for charge exchange $\pi^0$scattering reaction Eq.(\ref{pppp})
 have been considered in depth recently~\cite{Kaminski:2006qe}:
\begin{equation}
\sigma = \frac{16\pi}{9}\sqrt\frac{s-4M^2_{\pi^{\pm}}}{s-4M^2_{\pi^{0}}}
(a^{(0)}_0-a^{(2)}_0)^2; \label{pipipipi}
\end{equation}
where $a^{(0)}_0 - a^{(2)}_0 = 0.27/M_{\pi^{\pm}}$
This is the dominant process for charge pion production, subject to  presence of $\pi^0$.
\item
For process Eq.(\ref{ffpp}),  the cross section of $\pi^{\pm}$ production in photon fusion we use~\cite{Terazawa:1994at}:
\begin{equation}
\sigma_{\gamma\gamma \rightarrow \pi^+\pi^-} = \frac{2\pi \alpha^2}{s}\left(1-\frac{4m_{\pi}^2}{s}\right)^{1/2}
\left(\frac{m_V^4}{(1/2s + m_V^2)(1/4s+m_V^2)}\right), \label {ggpipi}
\end{equation}
where $m_V=1400.0$ MeV. As we will see from
numerical calculations given the cross sections for
$\gamma\gamma \rightarrow \pi^+\pi^-$ resulting production rates will be smaller than the charge exchange
$\pi^0\pi^0 \rightarrow \pi^+\pi^-$ reaction.
\item
For process Eq.(\ref{eepp}), the cross section of $\pi^{\pm}$ production in electron - positron fusion we use~\cite{Gounaris:1968mw}:
\begin{equation}
\sigma_{e^+e^-\rightarrow \pi^+\pi^-} = \frac{\pi \alpha^2}{3}\frac{(s-4m^2_{\pi})^{3/2}}{s^{5/2}}\left|F(s)\right|^2. \label{eepipi}
\end{equation}
The form factor $F(s)$ can be written in the form:
\begin{equation}
F(s) = \frac{m_{\rho}^2+m_{\rho} \Gamma_{\rho}d}{m_{\rho}^2-s+\Gamma_{\rho}(m_{\rho}^2/k_{\rho}^3)[k^2(h(s)-
h(m^2_{\rho}))+k_{\rho}^2h^{\prime}(m^2_{\rho})(m_{\rho}^2-s)]-im_{\rho}(k/k_{\rho})^3\Gamma_{\rho}(m_{\rho}/\sqrt{s})};
\end{equation}
where $h^\prime(s)=dh/ds$ and
\begin{equation*}
k=\left(\frac{1}{4}s-m_{\pi}^2\right)^{1/2};\quad
k_{\rho}=\left(\frac{1}{4}m_{\rho}^2-m_{\pi}^2\right)^{1/2};\quad
h(s) = \frac{2}{\pi}\frac{k}{\sqrt{s}}\ln\left(\frac{\sqrt{s}+2k}{2m_{\pi}}\right);
\end{equation*}
$m_{\rho}=775$ MeV, $\Gamma_{\rho}=130$ MeV, $d=0.48$. Given this cross section we also
find that the rate of charged pion production is  small when compared to $\pi_0$-charge exchange
scattering.
\item
For reaction (\ref{ggpi0pi0}) we have \cite{Mennessier:2007wk}:
\begin{equation}
\sigma(\gamma\gamma \rightarrow \pi^0\pi^0) =
\left(\frac{\alpha^2\sqrt{s-4m_{\pi}^2}}{8\pi^2\sqrt{s}}\right)\left[1+
\frac{m_{\pi}^2}{s}f_s\right]\sigma(\pi^+\pi^- \rightarrow \pi^0
\pi^0), \label{siggpi0}
\end{equation}
where
\begin{equation}
f_s =  2(\ln^2(z_+/z_-)-\pi^2)+\frac{m_{\pi}^2}{s}(\ln^2(z_+/z_-)+\pi^2)^2,
\end{equation}
and $z_{\pm} = (1/2)(1\pm \sqrt{s-4m_{\pi}^2})$.
\end{itemize}

\begin{figure}
\centering
\includegraphics[width=10.6cm,height=10cm]{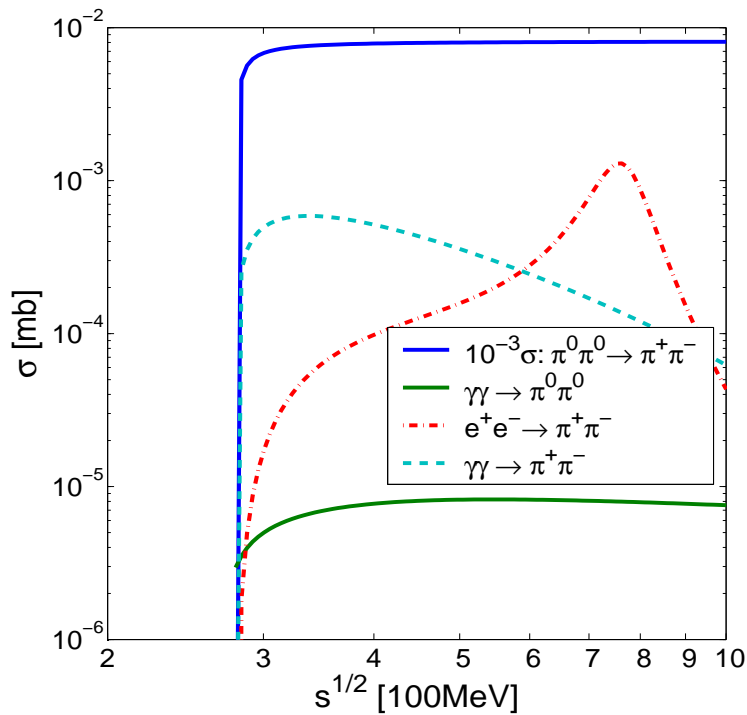}
\caption{\small{The cross section  $\sigma$ for
 pion  pair production, and pion charge exchange (solid top line),
as functions of $\sqrt{s}\le 1$ GeV$^2$.}}
\label{sigma}
\end{figure}

The cross sections for $\pi^+\pi^-$ pair  production, evaluated
using Eqs.(\ref{pipipipi}), (\ref{ggpipi}) and (\ref{eepipi}) are
presented in figure \ref{sigma} as functions of reaction energy
$\sqrt{s}$ . Top solid line (blue) is for charged pions production
in $\pi^0$ scattering Eq.(\ref{pppp}), the magnitude of this cross
section being very large we reduce it in presentation by factor
1000; the dashed line is for $\pi^+\pi^-$ production in photon
fusion Eq.(\ref{ffpp}); dash-doted line is for electron positron
fusion Eq.(\ref{eepp}). The bottom solid line (green) is for  photon
fusion into two neutral pions, Eq.(\ref{siggpi0}). The prediction
for $\sigma_{\gamma\gamma \rightarrow \pi^+\pi^-}$ is about 480 nb
(data 420 nb) at the peak near threshold \cite{Mennessier:2007wk},
which is in agreement with calculations presented here. The
reaction $\sigma_{\gamma\gamma \rightarrow \pi^0\pi^0}$(Eq.(\ref{ggpi0pi0}))
is much smaller than others and we do not consider this reaction further. We note
that some of these results are currently under intense theoretical
discussion as they relate to chiral symmetry. For our purposes the
level of precision of here presented reaction cross sections
is quite adequate.

\section{Numerical results}\label{resnum}

\subsection{Particle production relaxation times}
\begin{figure}
\centering
\includegraphics[width=10.6cm,height=14cm]{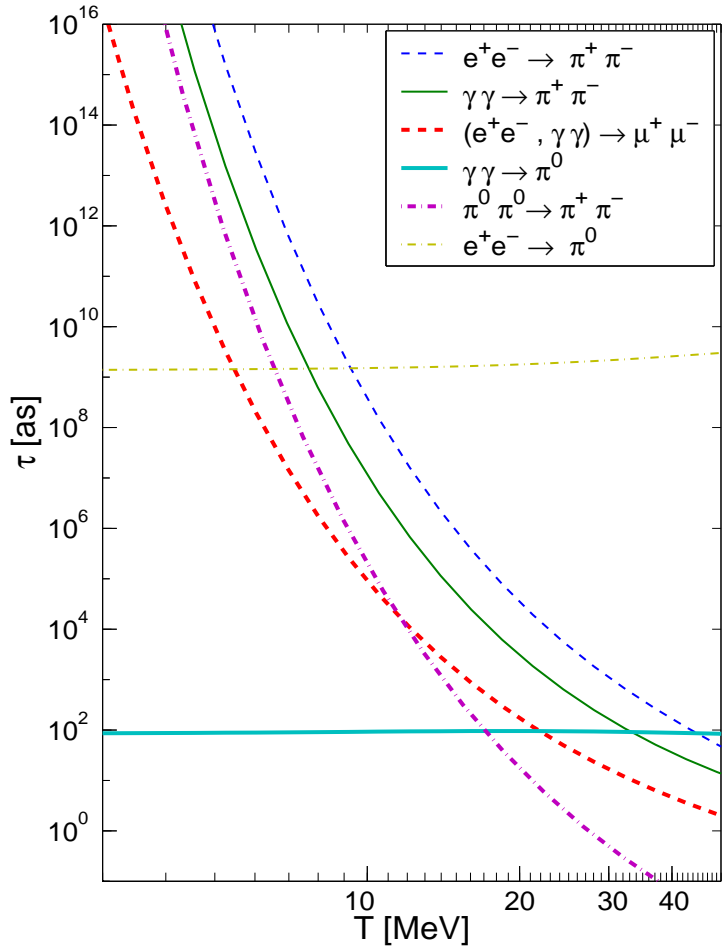}
\caption{\small{The relaxation time $\tau$ for  the different channels
of pion  and muon  production (see box), as functions of plasma
temperature $T$. }}
\label{taumupi}
\end{figure}

In figure \ref{taumupi} we show relaxation time $\tau$ for the different processes  considered as function of temperature $T\in [3,50]$ MeV.
Because of the  large difference in production rates which can be compensated by different densities of particles present
(magnitudes of fugacities) we introduce partial relaxation time for each of the three
 reactions $\pi^0\pi^0 \rightarrow \pi^+\pi^-$, $\gamma\gamma
\rightarrow \pi^+\pi^-$ and $e^++e^- \rightarrow \pi^+\pi^-$:
\begin{equation}\label{tausc}
\tau_{\pi^0\pi^0\leftrightarrow\pi^{+}\pi^{-}} = \frac{1}{2}\frac{{dn_{\pi^{\pm}}}/
{d\Upsilon_{\pi^{\pm}}}}{{R_{\pi^0\pi^0\leftrightarrow \pi^+\pi^-}}};\quad
\tau_{\gamma\gamma\leftrightarrow \pi^+\pi^-} =
    \frac{1}{2}\frac{{dn_{\pi^{\pm}}}/{d\Upsilon_{\pi^{\pm}}}}{R_{\gamma\gamma\leftrightarrow \pi^+\pi^-}}; \quad
\tau_{e^+e^-\leftrightarrow \pi^+\pi^-} =
    \frac{1}{2}\frac{{dn_{\pi^{\pm}}}/{d\Upsilon_{\pi^{\pm}}}}{R_{e^+e^-\leftrightarrow \pi^+\pi^-}}.
\end{equation}
When $T\ll m$, we can use the Boltzmann approximation to the particle distribution functions. Since in this limit
the density is proportional to  $\Upsilon$ the  relaxation times  doesn't
depend on   $\Upsilon$. Moreover, even for $T\to 50$ MeV, we have for muons $e^{-m/T}\simeq 1/3$, thus quantum
correlations in phase space remain small, and the Boltzmann limit can be employed.
To account for small deviation from Boltzmann limit  arising towards the upper limit
of the temperature range we consider,  that is at $T \simeq 50$ MeV,
we used the exact equations with  $\Upsilon_i = 1$  to calculate $\tau$ for each case, 
value corresponding to the maximum density that can be 
reached  for a given  temperature,   for which the quantum effect is largest.
In addition to these three cases Eq.(\ref{tausc})  we show in figure \ref{taumupi}
the muon production relaxation time Eq.(\ref{taumu} ),
the two photon fusion into $\pi^0$ relaxation time Eq.(\ref{taupi0}), a nearly horizontal line (turquoise, bottom),
which is slightly greater than the free space $\pi^0$ decay rate. Finally, the thin dash-dot line at about $10^8$ times
greater value of time is the electron-positron fusion into $\pi^0$, Eq.(\ref{pi0ee}).

\subsection{Rates of pion and muon formation}

\begin{figure}
\centering \hspace*{0.99cm}
\includegraphics[width=8.6cm,height=12.5cm]{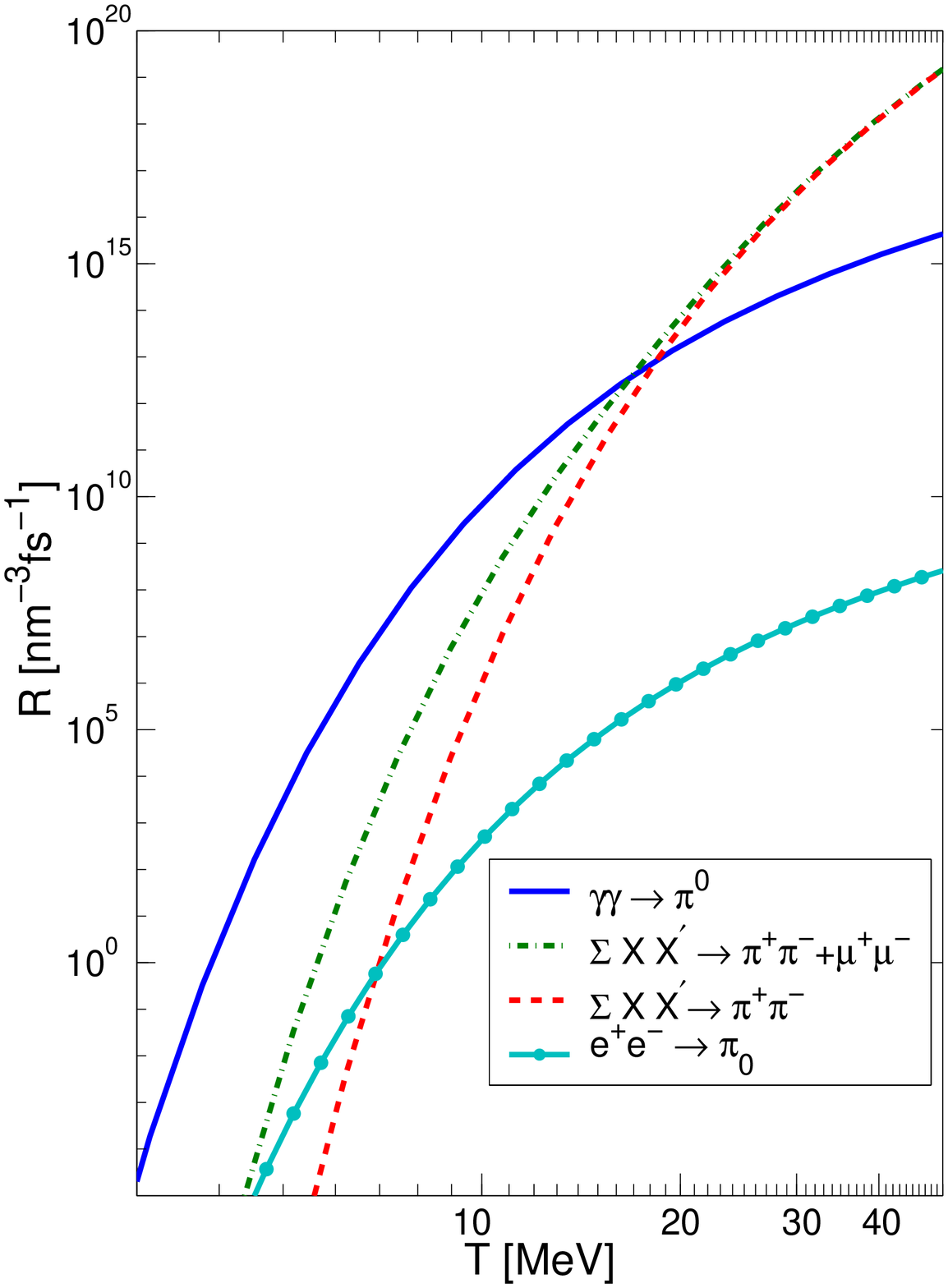}\hspace*{-0.69cm}
\includegraphics[width=8.6cm,height=12.5cm]{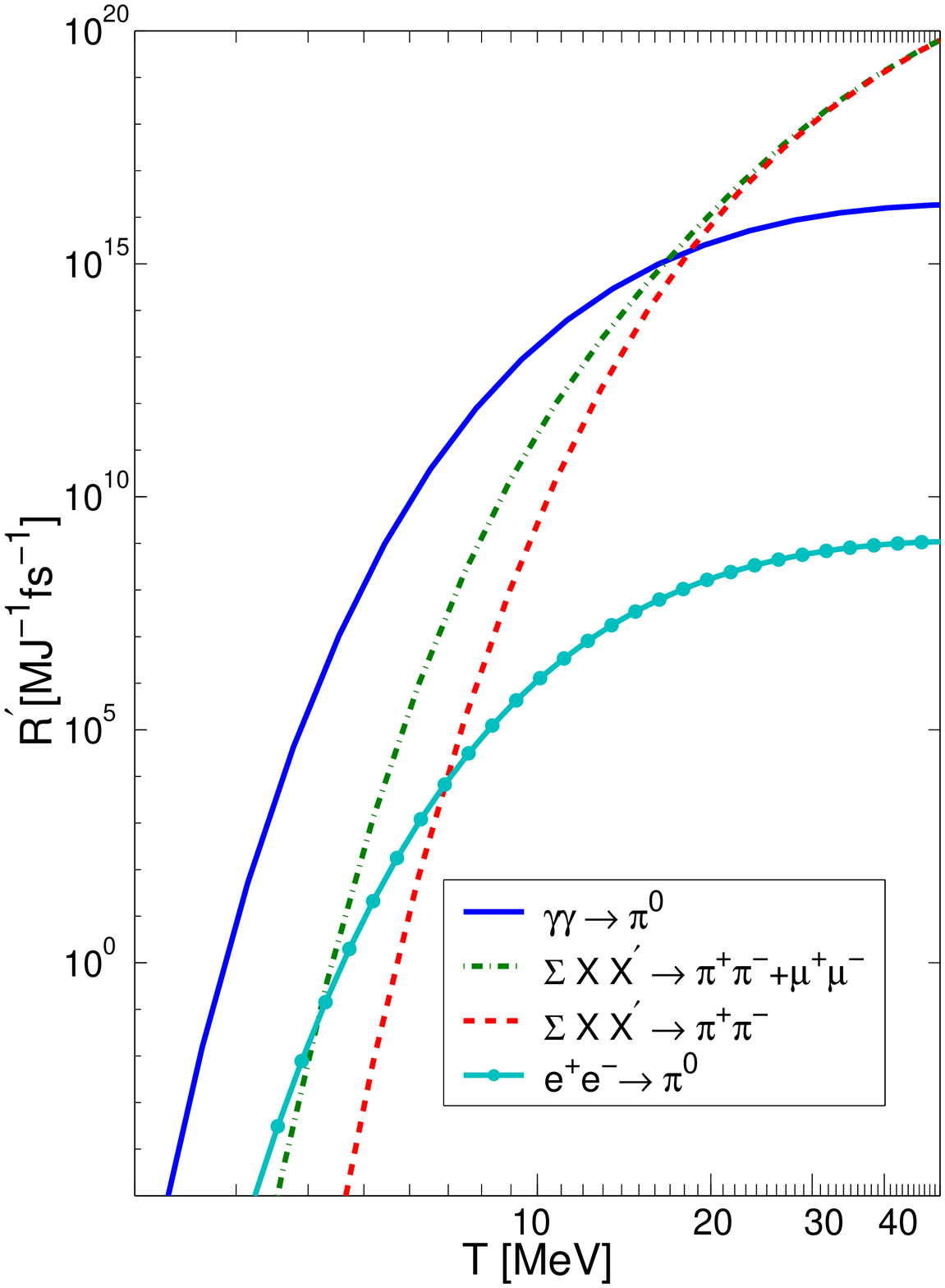}
\caption{\small{On left, the   invariant pion production rates in units of nm$^{-3}$fs$^{-1}$,
as a function of temperature $T$. On  right the production rate $R^\prime$ per  Joule energy content
in the fireball,  in units of MJ$^{-1}$fs$^{-1}$, in both cases for reactions shown in the box.}} \label{pions}
\end{figure}

In figure \ref{pions} we show on left as a solid (blue) line as a function of fireball temperature
 the rate per unit volume  and time for the process  $\gamma+\gamma\to \pi^0$, the dominant
mechanism of pion production. The other solid line with dots corresponds to $e^++e^-\to \pi^0$ reaction
which in essence remains, in comparison, insignificant. Its importance follows from the fact that it provides
the second most dominant path to $\pi_0$ formation at lowest temperatures considered, and it
operates even if and when photons are not confined to remain in the plasma drop.

We  improve   the rate presentation on the right hand side in figure \ref{pions}: considering that  the formation of a
plasma state involves an experimentally given fireball energy content $\cal E$ in Joules,
we use Eq.(\ref{SB})  to eliminate the volume $V$  at each temperature $T$:
\begin{equation}
R^{\prime}_{\pi^0}\equiv \frac {d^2 W^{\prime}_{\gamma\gamma\to \pi^0}}{dt d {\cal E} }
                      =\frac{1}{g\sigma T^4}\frac {d^4W_{\gamma\gamma\to \pi^0}}{  d V dt} = \frac{1}{g\sigma T^4}  R_{\pi^0}
\end{equation}
For chemical nonequilibrium, replace $g\to \Upsilon^2_\gamma g(\Upsilon)$.
Considering the (good) approximate Eq.(\ref{taupi01}) we  obtain:
\begin{equation}
R^{\prime}_{\pi^0} \simeq
      \left(\frac{m_{\pi} }{2\pi T}\right)^{3/2}\frac{e^{-m_{\pi}/T} }{ g\sigma T \tau_{\pi^0}^0 } .
\label{taupi02}
\end{equation}
We use units such that $\hbar=c=k=1$ and thus $R^{\prime}$  is a
dimensionless expression. Recalling the value of these constants, the  units we used for $R^{\prime}$ derive from
  MeV s=1.603\,10$^{-4}$ MJ fs.  

The other lines in figure \ref{pions}  address the sum of  formation rates of charged pion pairs (dashed, red) by all reactions
considered in this work, $\pi^{0}+\pi^{0}\to \pi^{+}+\pi^{-},  \gamma+\gamma \to \pi^{+}+\pi^{-}, e^++e^-\to \pi^{+}+\pi^{-}$.
We also present  the sum of all reactions leading to either a charged pion pair, or muon pair (dot-dashed, green) lines,
that is adding in .  $\gamma+\gamma \to \mu^{+}+\mu^{-}, e^++e^-\to \mu^{+}+\mu^{-}$.
The rationale for this presentation is that we  do not care how a heavy particle is produced, as long
as it can be observed. The dashed (red) line assumes that we specifically look for charged pions, and
dot-dashed (green) line that we wait till charged pions decays, being  interested in  the total final muon yield.
 The $\pi^0$ production
rate (blue, solid line) is calculated using Eq.(\ref{pi0pr}) and
yields on the logarithmic scale nearly indistinguishable result from
the approximation Eq.(\ref{taupi01}). For $\pi^\pm$ production we
refer to section \ref{pichprod} and for $\mu^\pm$ production we
refer to \ref{muprod}.

In table \ref{VTN} we show the values of key reaction rates $R$ and relaxation times $\tau$ at $T=5$ and $15$ MeV.
We note the extraordinarily fast rise of the rates with temperature, in some instances bridging 15 -- 20 orders in magnitude
when results for $T=5$ and $15$ MeV are compared.
\begin{table}
\caption{Values of  rates, relaxation times for all reactions at $T=5$ MeV and  $T=15$ MeV}
 \label{VTN}
\begin{tabular}{|c|c|c|c|c|c|}
  \hline
             &   $T=5$ MeV  & $T=5$ MeV &$T=15$ MeV & $T=15$ MeV\\
 reaction & $\tau$ [as] & $R$ $[\rm{nm^{-3}fs^{-1}}]$& $\tau$ [as] & $R$ $[\rm{nm^{-3}fs^{-1}}]$   \\
  \hline
$\gamma\gamma \leftrightarrow \pi_0$ & $88$ & $3.3\,10^3$ & $95$ & $1.2\,10^{12}$\\
$e^+e^- \leftrightarrow \mu^+\mu^-$ & $1.2\,10^{10}$ &  $3.2\,10^{-3}$ & $1.9\,10^3$ & $1.5\,10^{11}$ \\
$\gamma\gamma \leftrightarrow \mu^+\mu^-$ & $1.0\,10^{10}$ & $3.7\,10^{-3}$ &$1.3\,10^3$& $2.1\,10^{11}$ \\
$\pi^0\pi^0 \leftrightarrow \pi^+\pi^-$& $2.9\,10^{12}$ & $2.1\,10^{-8}$& $4.6\,10^2$ & $9.5\,10^{10}$  \\
$\gamma\gamma \leftrightarrow \pi^+\pi^-$& $6.4 \, 10^{13}$ & $9.7\,10^{-10}$ & $5.1\,10^4$ & $8.7\,10^8$   \\
$e^+e^- \leftrightarrow \pi^+\pi^-$& $7.8\,10^{15}$ & $7.9\,10^{-12}$ & $9.5\,10^5$ & $4.6\,10^{7}$\\
\hline
\end{tabular}
\end{table}

\begin{figure}
\centering\hspace*{0.99cm}
\includegraphics[width=8.6cm,height=11.5cm]{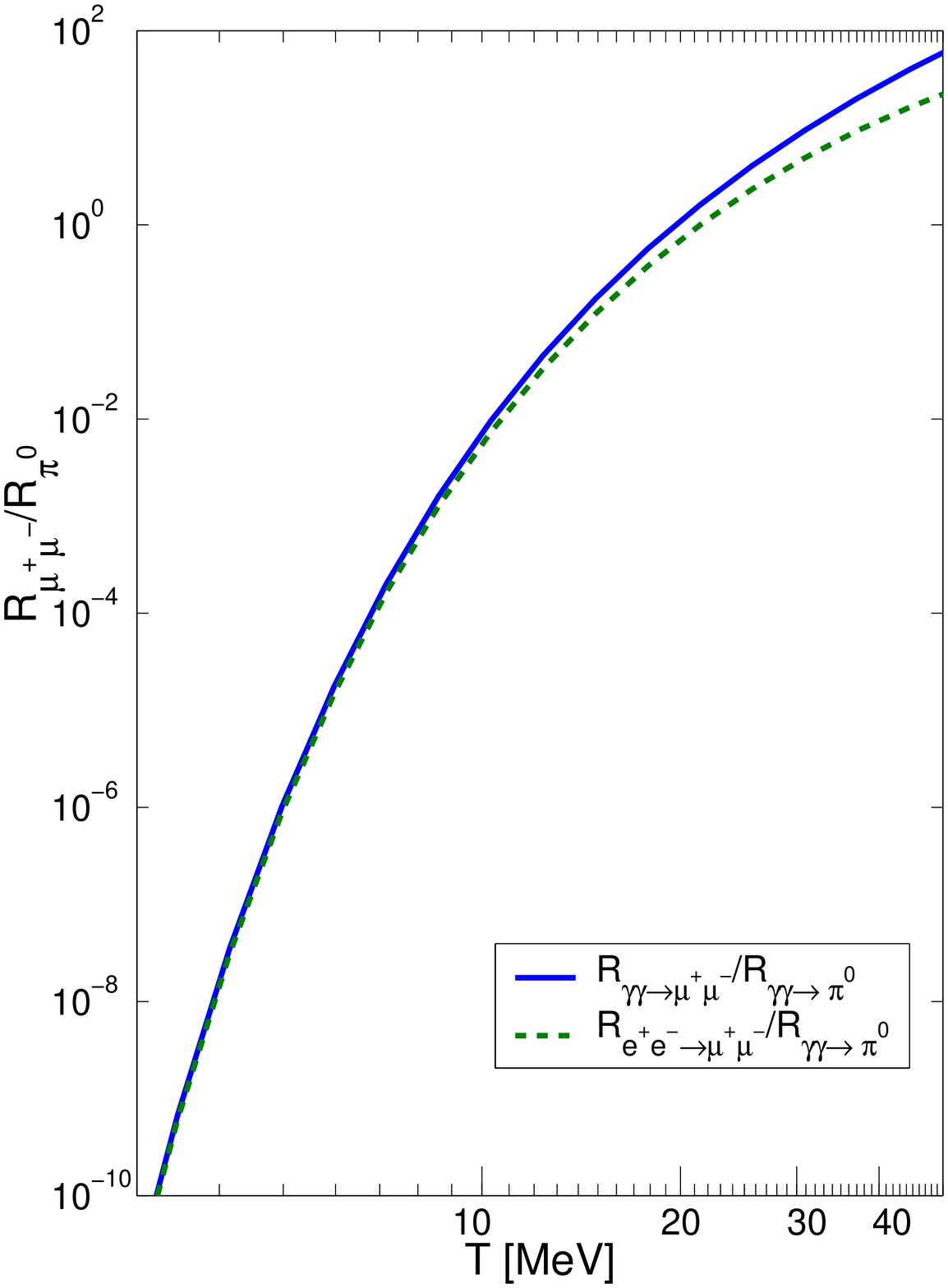}\hspace*{-0.7cm}
\includegraphics[width=8.6cm,height=11.5cm]{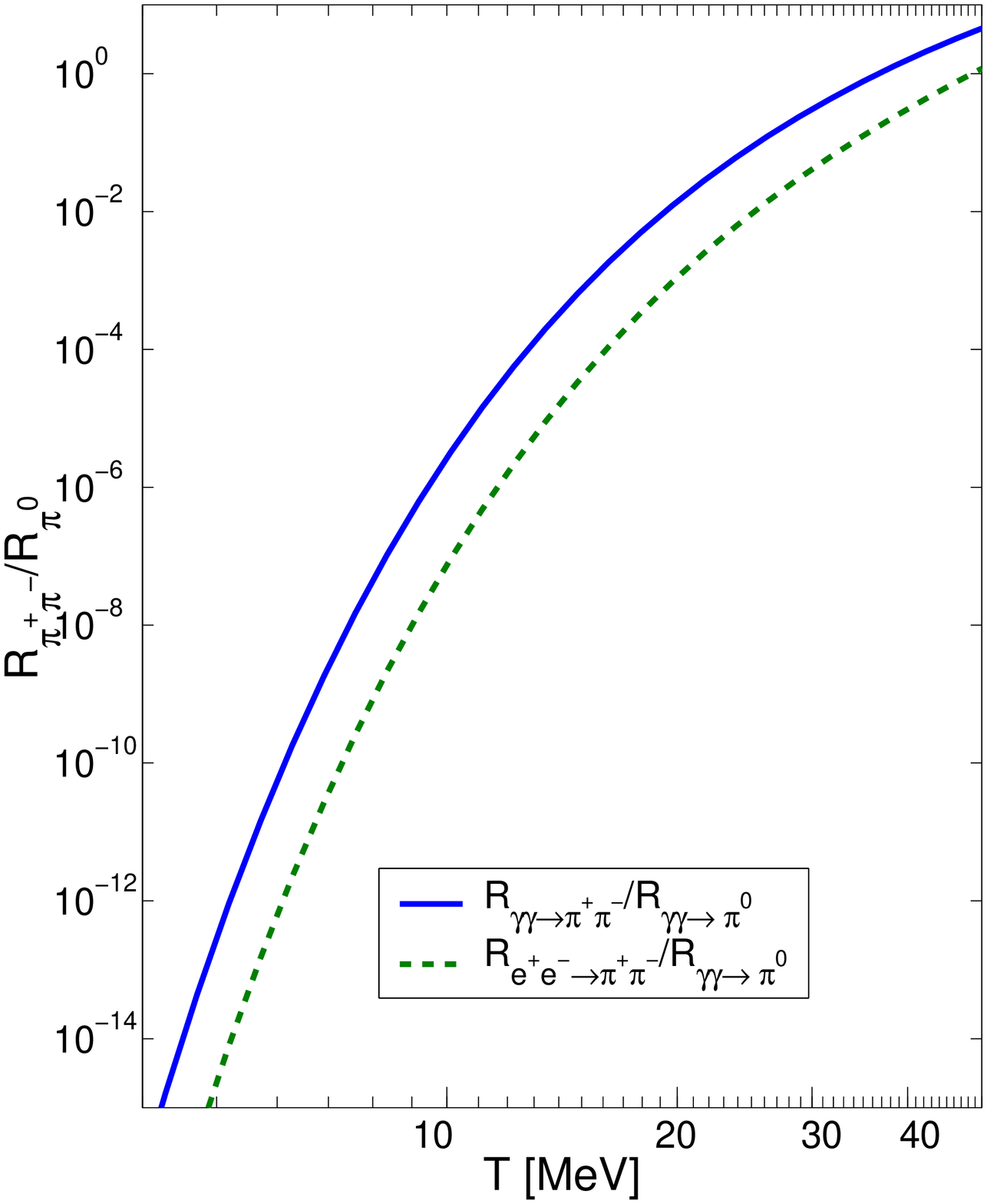}
\caption{\small{On left: Muon and on right charged pion production rates in electromagnetic processes
normalized by  $\pi^0$ production rate. Solid line (blue) for  $\gamma\gamma$ , dashed line  (green) for
 $e^+e^- $ induced process.}}
\label{mupir}
\end{figure}

In order to understand the individual contributions to the  different reactions
entering the sum of rates presented above, we show  as a function of temperature
in the figure \ref{mupir}   the relative strength of muon pair (left)
and charge pion (on right) electromagnetic ($\gamma+\gamma, e^++e^-$)
 production,  using as the reference the
 $\gamma+\gamma \rightarrow \pi^0$ reaction.
The $\mu^{\pm}$ production rates are calculated using
Eq.(\ref{ratemu}) with $|M|^2$ from Eq.(\ref{m2emu}) and
 Eq.(\ref{m2gmu}) respectively. This ratio is smaller than unity for $T\lessim 20$ MeV.
For larger $T$, the  muon direct production rate becomes
larger than $\pi^0$ production rate. Charged pions (on right in figure  \ref{mupir})
can be produced in direct reaction at a rate larger than neutral pions
only for $T>35$ MeV. The photon channel dominates.

\section{Discussion and Conclusions}\label{concl}

We found that the production of $\pi^0$ is the dominant coupling of electromagnetic radiation
to heavy (hadronic) particles with $m \gg T$,  and as we have here demonstrated
 that noticeable particle yields can be expected already
at modest temperatures $T\in [3,10]$ MeV.
In present day  environment of 0.1 --1 J plasma  lasting a few fs, our results suggest that we can expect
integrated over space-time evolution of the EP$^3$ fireball a  $\pi^0$  yield at the limit of detectability.
For $T \to 15$ MeV the $\pi^0$ production rate  remains dominant and indeed
very large, reaching the production rate $R^{\prime}\simeq 10^{15}$[MJ$^{-1}$fs$^{-1}]$.
Charge exchange reactions convert some of the neutral pions into charged pions which are
more easy to detect.

In this situation it is realistic to consider the possibility of forming a chemically equilibrated fireball
with $\pi^0,\pi^\pm,\mu^\pm$ in chemical abundance equilibrium. The heavy particles are produced
in early stages when temperature reached is highest. Their abundance in the fireball  follows the fireball
expansion and cooling till their
freeze-out, that is decoupling of population equation production rates. The particle yields are than
given by the freeze-out conditions, specifically the chemical freeze-out temperature $T_f$ and
volume $V_f$, rather than the integral over the rate of production. In this situation the heavy
particle yields become diagnostic tools of the freeze-out conditions, with the mechanisms of
their formation being less accessible. However, one can avoid this condition by appropriate
staging of fireball properties.

The present study has not covered, especially for low temperature range all the possible
mechanisms, and we addressed some of these issues in the introduction. Here we note further
that the production of heavy particles
requires energies of the magnitude $m/2$ and thus   is due to collisions involving   the  (relatively speaking)
far tails of a thermal particle distribution. If these tails  fall off as a power law, instead of the
Boltzmann exponential  decay~\cite{Biro:2004qg},  a  much greater yield of heavy particles
 could ensue. There could further be present a   collective amplification
to the production process e.g. by residual matter flows, capable to
enhance the low temperature yields, or by collective plasma oscillations and inhomogeneities.

These are just some examples of many
reasons to hope and  expect a   greater particle yield than we
computed here in microscopic and controllable two particle reaction
approach. This consideration, and our encouraging
`conventional' results suggest that
 the study of  $\pi^0$ formation in  QED plasma  is of considerable  intrinsic interest. Our results
provide a lower limit for rate of particle production and when folded with models of EP$^3$ fireball
formation and evolution, final yield.

It is of some interest to note that  the study of pions in QED plasma allows exploration of pion properties
in electromagnetic medium. Specifically,    recall
 that  1.2\% fraction of  $\pi^0\to e^+e^-\gamma$ decays, which
implies that the associated processes such as  $e^++e^-\to \gamma +\pi_0$ are  important. We cannot
evaluate this process at present as it involves significant challenges in understanding of $\pi_0$
off-mass shell `anomalous' couping to two photons.

The experimental environment we considered here should allow
a detailed study of the properties of   pions (and also muons) in a thermal background.
There is considerable fundamental
interest in the study of   pion properties and specifically
pion mass splitting  in  QED plasma at temperature $T\gtrsim \Delta m$ and in presence of
electromagnetic fields. We already have shown that due to quantum statistics effects, the effective
in medium decay width of  $\pi^0$  differs  from the free  space value, see figure \ref{taupi0app}.
In addition,   modification of mass and  decay width due to ambient medium influence on
the pion internal structure is to be expected.
 Further we hope that the study of pions in the  EP$^3$ fireball  will contribute to the better
understanding of   the relatively large difference in mass between
 $\pi^0$and $\pi^\pm$. The  relatively large size of the PE$^3$ environment should 
make  such changes, albeit small, measurable.

The experimental study of  $\pi^0$ in QED plasma environment is not
an easy task. Normally, one would think that the study of the
  $\pi^0$ decay into two  67.5 MeV $\gamma$ (+ thermal Doppler shift motion)
produces a  characteristic signature. However, the   $\pi^0$ decay is
in time and also in location overlapping with the plasma formation and disintegration. The debris of the plasma,
 reaches any detection system  at practically the same time instance as does the 67.5 MeV $\gamma$. The large
amount of available radiation will disable  the detectors. On the other hand we realize that
the hard thermal component of the plasma, which leads to the production of   $\pi^0$ in the early fireball stage, is
most attenuated by plasma dynamical expansion.  Thus it seems possible to plan for  the  detection of
 $\pi^0$ e.g. in a heavily shielded detection system.

The decay time of charged pions being 26 ns, and that of charged muons being 2.2 $\mu$s
it is possible to separate in time the plasma debris from the decay signal of these particles.
Clearly, these heavy charged particles can be detected with much greater ease, also considering that
the decay product of interest is charged.
For this reason, we also have in depth considered all channels of production of charged pions
and muons. Noting that practically all charged pions turn into muons, we  have also compared
the production rates of $\pi^0$ with all heavy particles, see dot-dashed (green) line in figure \ref{mupir}.
This comparison suggests that for plasmas at a temperature reaching $T>10$ MeV the production
of final state muons will most probably be by far easier to detect. On the other hand for $T<5$ MeV
it would seem that the yield difference in favor of  $\pi^0$ outweighs the detection system/efficiency loss
considerations. Future work addressing non-conventional processes will show at how low  $T$
we can still expect observable heavy particle yields.

An   effort to detect $\pi^0$ directly is justified since   we can learn  about the
properties of the plasma (lifespan, volume and temperature in early stages) e.g.
from a comparative study of  the $\pi^0$ and  $\pi^\pm$ production.
We have found that  at about $T>16$ MeV, the pion charge exchange  $\pi^0\pi^0\to \pi^+\pi^-$
reaction for chemically equilibrated $\pi^0$ yield is  faster than the natural $\pi^0$ decay,
 and the chemical equilibration time constant,  see the dot-dashed line  in
figure \ref{taumupi}. Thus beyond this temperature the yield of charged pions can be expected to be
in/near chemical equilibrium for a plasma which lives at, or above this temperature, for longer than 100 as.

In such an environment the yield of $\pi^0$ is expected to be near chemical equilibrium, since the
decay rate is compensated by the production rate, and, within 100 as, the chemical
equilibrium yield is attained. Moreover,
the thermal speed of produced $\pi$ can be  obtained from the nonrelativistic
relation $\frac 12 m \langle v^2\rangle =\frac 3 2 T$,  thus $\overline v \propto  \sqrt T $ and, for $T=10 $ MeV,
$\overline v\simeq 0.5 $c. This is
nearly equal to the sound velocity of EP$^3$,  $v_s\simeq c/\sqrt{3}=0.58c$. Thus the heavy $\pi^0$ particles
can be seen as co-moving with the expanding/exploding  EP$^3$, which completes
the argument to justify their transient chemical
equilibrium yield in this condition.

The global production yield of neutral and charged pions
should  thus  allow the study of volume and temperature history of the QED  plasma.
More specifically, since with decreasing temperature, for $T<16$ MeV, there is a rapid  increase of the relaxation time for
 the charge exchange  process, there is a rather rapid drop
 of the  charged pion yield below chemical equilibrium --- we note that charge exchange   equilibration time at $T=10$ MeV is
a factor $10^5$ longer. We note that the study of two pion correlations provides an independent measure of
the source properties (HBT measurement).

The relaxation time of electromagnetic production of muon pairs wins over $\pi^0$
relaxation time for $T>22$ MeV, see dashed line, red, in figure \ref{taumupi},
the  direct  electromagnetic processes
of charged pion production (thin green, solid line for $ \gamma\gamma \rightarrow
\pi^+\pi^-$ and dashed, blue for $e^+e^- \rightarrow \pi^+\pi^-$) remain  sub-dominant.
Thus for $T>22$ MeV we expect, following the same chain of arguments for muons as above for
charged pions, a near chemical equilibrium yield. If the study of all these
$\pi^0,\pi^\pm,\mu^\pm$ yields,  their spectra and even pion correlations were possible, considerable insight into
 $ e^-,   e^+, \gamma $ plasma (EP$^3$) plasma formation and dynamics  at $T<25$ MeV can be achieved.

\subsubsection*{Acknowledgments}
This research was supported by the DFG Cluster of Excellence:
Munich Center for Advanced Photonics and
by a grant from: the U.S. Department of Energy  DE-FG02-04ER4131.

\vspace*{-0.3cm}

\end{document}